%% file: main.tex
\documentclass{article}
\usepackage[a4paper,margin=3.8cm]{geometry}

\usepackage{graphicx} 

\usepackage[utf8]{inputenc} 
\usepackage[T1]{fontenc}    
\usepackage{hyperref}       
\usepackage{url}            
\usepackage{natbib}
\usepackage{authblk} 
\usepackage{xcolor}
\usepackage{xspace}
\usepackage{array}

\newcommand{\leo}{LEONARDO\xspace}
\newcommand{\ib}{InfiniBand\xspace}
\newcommand{\gbit}{Gbps\xspace}
\newcommand{\nvidia}{NVIDIA\xspace}
\newcommand{\intel}{Intel\xspace}
\newcommand{\flops}{FLOPS\xspace}
\newcommand{\pf}{peta\flops}
\newcommand{\gpuname}{\emph{Da Vinci} A100\xspace}
\newcommand{\davinci}{\emph{Da Vinci}\xspace}
\newcommand{\tf}{tera\flops}
\newcommand{\gf}{giga\flops}

\usepackage{multirow}   
\newcommand\T{\rule{0pt}{2.6ex}}       
\newcommand\B{\rule[-1.2ex]{0pt}{0pt}} 

\newcommand*{\email}[1]{\normalsize\href{mailto:#1}{#1}\par}

\providecommand{\keywords}[1]
{
  \small	
  \textbf{\textit{Keywords---}} #1
}

\hypersetup{
    colorlinks,
    linkcolor={red!40!black},
    citecolor={blue!50!black},
    urlcolor={blue!80!black},
    pdftitle={\leo: A Pan-European Pre-Exascale Supercomputer for HPC and AI Applications},
    pdfsubject={cs.AR, cs.DC},
    pdfauthor={Matteo Turisini, Giorgio Amati, Mirko Cestari},
}

\title{\leo: A Pan-European Pre-Exascale Supercomputer for HPC and AI Applications}

\author[a,*]{Matteo Turisini} 
\author[b]{Giorgio Amati}
\author[a]{Mirko Cestari}

\affil[a]{SuperComputing Applications and Innovation Department, CINECA, Via Magnanelli 6/3, Bologna, 40033, Italy}
\affil[b]{SuperComputing Applications and Innovation Department, CINECA, Via dei Tizii, 6, Rome 00185, Italy}
\affil[*]{Corresponding author: Matteo Turisini \email{m.turisini@cineca.it}}

\begin{document}

\maketitle
\input{abstract.tex}

\input{intro.tex}

\input{bridge.tex}

\input{booster.tex}

\input{network.tex}
\input{storage.tex}

\input{service.tex}
\input{software.tex}
\input{power.tex}

\input{outro.tex}

\newpage
\bibliographystyle{apalike}
\bibliography{references.bib}  

\newpage
\appendix
\input{benchmarks.tex}

\newpage
\input{hardware.tex}

\end{document}

%% file: abstract.tex
\begin{abstract}
A new pre-exascale computer cluster has been designed to foster scientific progress and competitive innovation across European research systems, it is called \leo.
This paper describes the general architecture of the system and focuses on the technologies adopted for its GPU-accelerated partition.
High density processing elements, fast data movement capabilities and mature software stack collections allow the machine to run intensive workloads in a flexible and scalable way.
Scientific applications from traditional High Performance Computing (HPC) as well as emerging Artificial Intelligence (AI) domains can benefit from this large apparatus in terms of time and energy to solution.
\end{abstract}

%% file: intro.tex
\section{Introduction} \label{sec:intro}

\leo is a new European computer cluster with pre-exascale computing capabiliy, at the level of $0.2 \times 10^{18}$~floating point operations per second (\flops).
The project has been conceived by \leo \emph{Consortium}, a group of six signatory countries\footnote{The countries are: Italy (project leader), Austria, Greece, Hungary, Slovakia and Slovenia.} of the European declaration on High Performance Computing~\citep{Ref:euroDeclarationHPC} whose purpose is to foster scientific and technological federative innovation across the European Union.
\leo is owned by the European High Performance Computing Joint Undertaking initiative~\citep{Ref:JU} and is hosted by CINECA interuniversity consortium~\citep{ref:cineca} at the Tecnopolo Manifattura Data Valley Hub in Bologna, Italy~\citep{Ref:tecnolopolo}.

The foreseen operational lifetime of the machine is 5 years.
In this period it is going to serve as a research facility for a broad class of scientific investigations, due to a complete set of state-of-the art hardware and software technologies that are presented in this paper.
The most relevant are a massive amount of computational power available at single node (i.e. a peak performance of 78 \tf), a fast access storage (over a TB/s bandwidth) and a flexible scalability for multi-node computations.
With \leo, researchers from academia and industry can tackle many challenges in different crucial fields, like Digital Twins applications, e.g. \cite{Ref:DTGEO}, Data-driven projects, e.g. \cite{Ref:GEOIN} and Urgent-Computing, e.g.~\cite{Ref:CHEESE2} to name a few.

\begin{figure} 
    \centering
    \includegraphics[width=\textwidth]{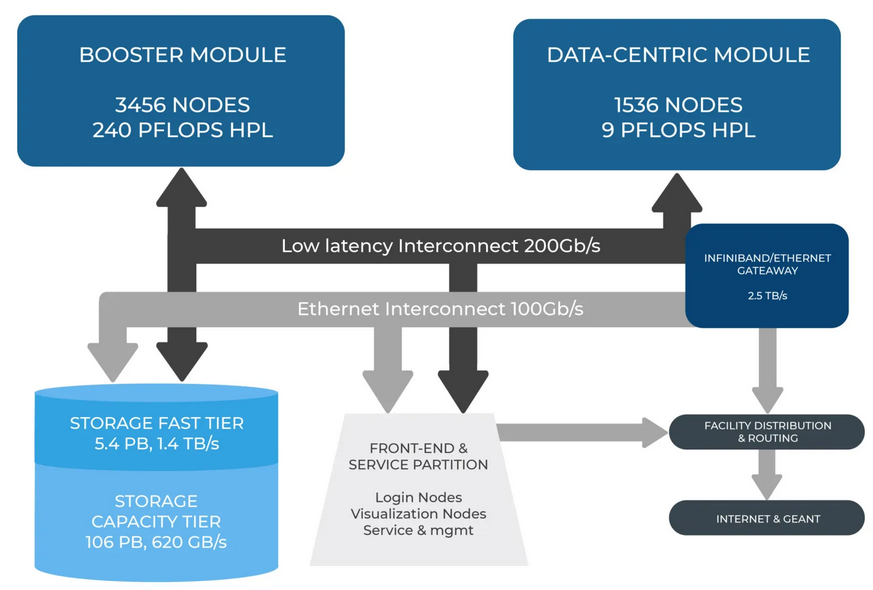} 
    \caption{Architectural overview}
    \label{fig:Schema}    
\end{figure}

The supercomputer has been designed by ATOS company with its technology partners and is composed by two compute partitions that are coupled with interconnects, storage and service subsystems.
A general purpose Data-Centric (DC) partition is intended to fulfill a vast range of traditional HPC applications by leveraging the latest central processing unit (CPU) technologies. 
It measures 1536 compute nodes, based on the \intel's 4th generation of Xeon Scalable processor, codenamed \emph{SapphireRapids}.
The CPU model is the 56-core $8480$\texttt{+} that features several hardware accelerators to support Single Instruction Multiple Data extension (SIMD) on top of the x86 instruction set.
Accelerated functionalities include cryptography and vector algebra \citep{Ref:SPR}.
The other compute partition is a heterogeneous module called Booster which is dedicated to applications that can benefit from the parallelism offered by general purpose graphical processing unit (GPU). 
The Booster consists in 3456 nodes configured with a single-socket host \intel \emph{Ice Lake} CPU ~\citep{Ref:8358} and four NVIDIA A100 \emph{Tensor Core} GPU chips~\citep{Ref:A100}.
The internal network, used for inter-node communication, relies on 200~\gbit Mellanox's \ib High Data Rate (HDR) technologies~\citep{Ref:HDR} and is organized in a \emph{dragonfly+} topology.
The storage is composed of a mix of high-speed and high-capacity appliances to accommodate  the requirements of modern Big Data and AI applications, including Cloud services and Interactive computing.
The infrastructure is completed by an operational 100~\gbit Ethernet network, 11 management nodes and 32~frontend servers where users can land, develop codes, submit jobs, and analyze results.
Figure~\ref{fig:Schema} presents a schematic overview of \leo. 
All subsystems are shared between the two compute partitions.
A set of four Ethernet/\ib gateways allows the cluster to be connected to external networks.

This paper describes the overall architecture of \leo and focuses on the Booster module.
Section \ref{Ref:Booster} presents Booster's node including computing elements and organization.
Details on network and storage partitions can be found in \ref{sec:network} and \ref{sec:storage}.
This is followed by paragraph \ref{sec:service} on frontend and service resources.
Software tools and libraries are listed in \ref{L_SW}.
Finally the power supply and the cooling systems are presented in \ref{L_power}.
Additionally, some benchmark results are reported in Appendix~\ref{appendix:benchmark} and the list of hardware components can be found in Appendix~\ref{sec:hardware}.
The DC module will be detailed in a separate article.

%% file: bridge.tex
\section{System details} \label{sec:system}

\leo is a quite large apparatus consisting of 155 racks, 2 tons each.
The compute partitions are made up of 138 racks based on the ATOS BullSequana XH2000 cabinet, a platform that offers high level integration density and Direct Liquid Cooling capabilities~\citep{Ref:BullSequana}.
Table~\ref{Tab:Racks} shows how compute racks are organized in cells and how blade servers and node units compose each rack. 
One cell encompasses both Booster and DC type nodes and is called \emph{Hybrid} cell. 
An additional cell (the twenty-third) houses storage and service equipment.
This includes 12 racks equipped with DDN's appliances and 5 further ATOS racks dedicated to management and frontend servers.

\input{tableRack.tex}

%% file: tableRack.tex
\begin{table} 
\centering
\begin{tabular}{|c|c|c|c|c|c|c|c|} \hline                        
Type       & Cell & $\frac{Rack}{Cell}$  & $\frac{Blade}{Rack}$ &$\frac{Nodes}{Blade}$ & Rack  & CPU nodes & GPU nodes \T\B  \\ \hline
Booster    & 19   & 6     & 30    & 1         & 114   & -       & 3420 \T \B\\  \hline
DC         & 2    & 8     & 26    & 3         & 16    & 1248    & - \T \B\\  \hline

\multirow{2}{*}{Hybrid} & \multirow{2}{*}{1}    
                   & 2    & 18    & 1         & 2     &  -      & 36  \T\B\\ \cline{3-8} 
            &      & 6    & 16    & 3         & 6     & 288     & - \T\B\\

\hline 
\hline
Total      & 22   &     -   & -   &    -        &  138 & 1536 & 3456 \T\B \\
\hline

\end{tabular}
\caption{Compute partitions racks}
\label{Tab:Racks}
\end{table}

%% file: booster.tex
\subsection{Booster partition} \label{Ref:Booster}
The Booster is the first \leo's compute partition to go in full production in 3Q 2023.
It consists of 3456 heterogeneous nodes designed to create a significant speedup in traditional HPC and new AI applications.
In facts, this supercomputer is one of the top level facility in the world to support scientific investigation in many fields: with $238.7$ \pf of sustained Linpack performance, it reached the 4th spot in the TOP500 ranking in June 2023 \citep{Ref:Top500_2306}, being at the same time the largest supercomputer based on NVIDIA Ampere architecture, with about 14k GPUs.
However, the improvement brought by \leo is not only a matter of pure performance, instead, the design of the machine has been intended to accompany the evolution of computing architectures towards hardware specializations and to extend the support for workloads related with the training and the usage of large AI models.

\input{gpu_chip.tex}

\input{gpu_node.tex}

%% file: gpu_chip.tex
\subsubsection{GPU accelerator device} \label{subsec:gpu}

The A100 \emph{Tensor Core} GPU is an accelerator device introduced by \nvidia in 2020 based on the \emph{Ampere} micro-architecture~\citep{Ref:A100}.
In the fast changing accelerators market, it represented a breakthrough in terms of flexibility, computational power ($+24\%$ floating point, FP) and communication speed ($+73\%$ memory bandwidth) in comparison to its predecessor, the V100 \emph{Tensor Core} GPU based on the \emph{Volta} architecture ~\citep{Ref:V100}.

The \emph{Ampere} offers an upgraded compute unit structure (third-generation Tensor Core, TC) that extends hardware support for tensor math to a wider set of datatypes including both floating point and integer numerical formats.
Concerning floating point computation in double precision (FP64), the device offers an impressive peak performance of about 20~\tf for tensor operations and around 10~\tf for non-tensor math.
Together with the single precision performance treated below, this is particularly engaging for HPC communities that rely on very high precision representations in their models.
In fact, moving from V100 to A100, a speedup between x1.5 and x2.1 has been measured in HPC benchmarks spanning from molecular dynamics to geo-sciences \citep{ref:nvidia_arch_in_depth}.
On the AI side, a new numerical format called \emph{Tensor Float 32} (TF32) definitively enables the use of TC to accelerate the training of a vast number of neural network models.
The TF32 is a custom floating point format with 8-bit range (as in FP32) and 10-bit precision (as in FP16).
The halved precision does not affect the accuracy of the computations in the AI context and brings a significant speed up instead.
The FP32 data path has been kept for I/O operations and the TF32 is the default choice for computation, so the speedup benefit is transparent to the user (no code change).
For maximum speed in training, the supported tensor math includes the standard FP16 datatype (inherited from the previous generation TC) and the new AI dedicated BF16 datatype (8-bit range, 7-bit precision) which allow a factor x2 in throughput respect to TF32 and a factor x20 compared to non-tensor operations.
Integer arithmetic is supported as well, for example 8-bit operations have a peak performance of 624 teraOPS and INT4 and binary even more.

\input{tableGPU.tex}

Table~\ref{Tab:A200} displays the main specifications of the two generations (\emph{Ampere} and \emph{Volta})
and present the characteristic of the \gpuname variant installed in \leo.
The latter is a \emph{custom} model consisting in a $97\%$ implementation of the full A100 GPU design (124 vs 128 Streaming Multiprocessors, SM), while the \emph{standard} A100 uses $84\%$ of it (104 SM).

In addition, the A100 offers an instructions set called \emph{Sparse Tensor Core} that double the TC performance reported in Table~\ref{Tab:A200} when working with AI applications.
With this approach, which is referred to as \emph{Structural Sparsity} by the vendor ~\citep{ref:nvidia_ga100_whitepaper}, the pruning of the weights matrix is structurally constrained by zeroing two elements out of four in a row.
At inference time, an efficient use of hardware resources allows to gain a clean factor two in throughput.

%% file: tableGPU.tex
\begin{table} 
        \centering
        \begin{tabular}{|r|c|c|c|}
\hline
                               & Ampere A100 (custom) & Ampere A100   & Volta V100  \T \B \\
\hline
        FP64 [\tf]        & 11.2        & 9.7      & 7.8\T\\
        FP32 [\tf]        & 22.4        & 19.5     & 15.7\B\\
        \hline
        FP64 TC [\tf]     & 22.4        & 19.5     & n.a.\T\\
	TF32 TC [\tf]     & 179         & 156      & n.a.\\
 	FP16 TC [\tf]     & 358         & 312      & n.a.\\ 
 	INT8 TC [teraOPS] & 716         & 624      & n.a.\\
 	INT4 TC [teraOPS] & 1432        & 1248     & n.a.\B\\
\hline
        SM [\#]                &  124        &  108     &   80\T\\
        CUDA FP64 core [\#]    & 3968        & 3456     & 2560\\     
        CUDA FP32 core [\#]    & 7936        & 6912     & 5120\\
        CUDA Tensor core [\#]  &  496        &  432     &  640\B\\
        \hline
        Max Clock [MHz]        & 1395        & 1410     & 1530\T\\
        L2 Cache [MB]          &   32        &   40     &    6\\
        Memory  [GB]           &   64        &   40     &   16\\
        Memory BW  [GB/s]      & 1640        & 1555     &  900\\
        TDP     [W]            &  440        &  400     &  300\B\\        
        \hline
        \end{tabular}
	\caption{Comparison of GPU chips specifications and peak performance.}
	\label{Tab:A200}
\end{table}

%% file: gpu_node.tex
\subsubsection{GPU blade}
\begin{figure}
    \centering
        \includegraphics[width=\linewidth]{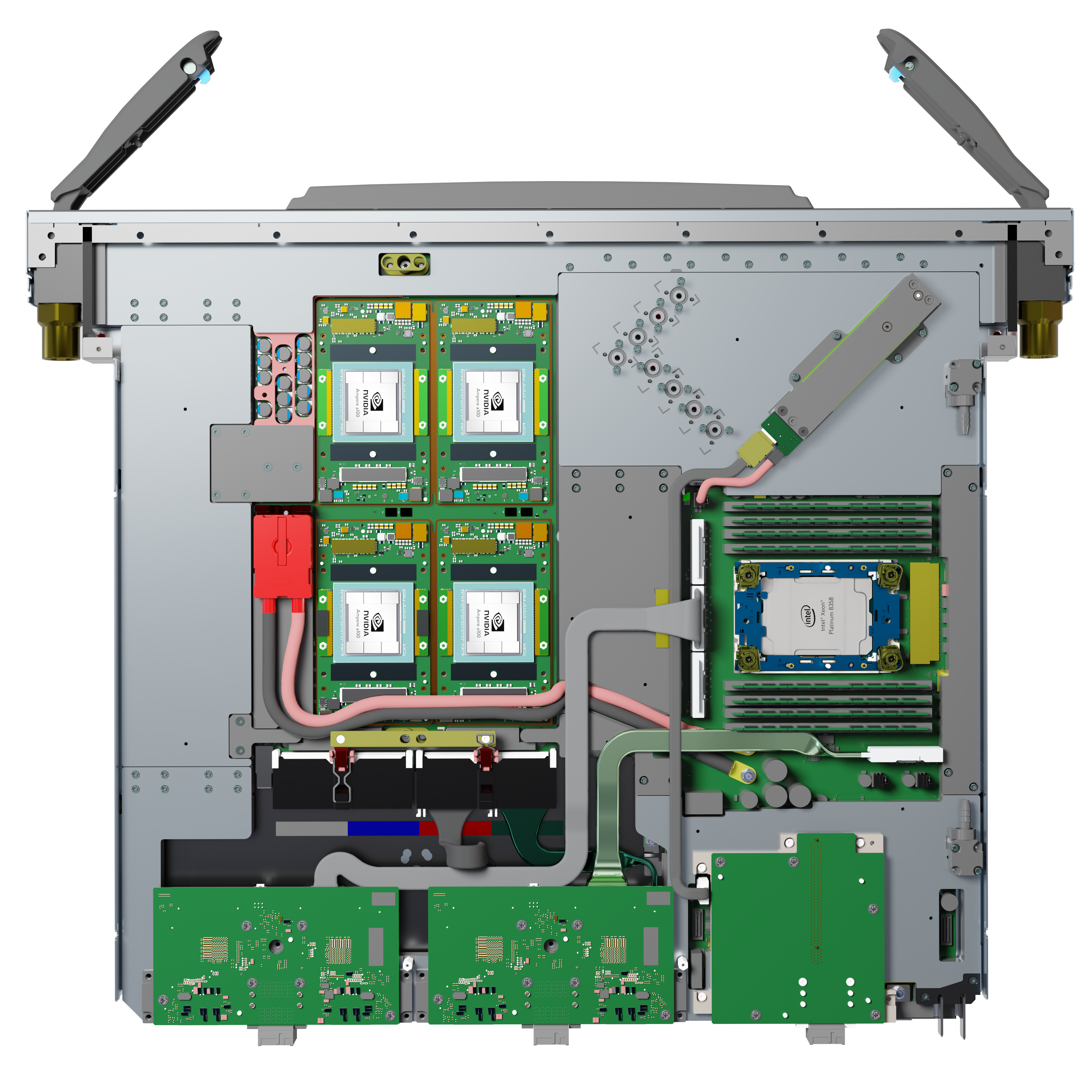}        
    \caption{GPU blade top view}
    \label{Fig:DaVinci_blade}
\end{figure}

The Booster's blade is a single node blade, based on the latest high-end GPU server board by ATOS company (BullSequana X2135).
The blade is called \davinci and a picture of it is shown in Figure~\ref{Fig:DaVinci_blade}.
The entire blade is liquid-cooled, so there are no fans onboard.

The host processor is a single socket \intel Xeon Platinum 8385 CPU~\citep{Ref:8358} with 32~cores and 48~MB cache (codenamed \emph{Ice Lake}).
The IceLake CPU is AVX-512 capable.
Each core contains two AVX-512 execution units which results in a 1024 operations per clock cycle and a peak performance of 2.6 \tf per core at the nominal frequency of 2.6 GHz. 
The memory subsystem is DDR4, clocked at 3200~MHz (6400~MT/s).
There are eight 32-bit memory controllers.
Each is capable of 25 GB/s for a total maximum bandwidth of 200GB/s for CPU-RAM communication.
The corresponding eight DIMM slots are equipped with 64~GB capacity modules, so the total RAM available on node is 512~GB.

Four \gpuname GPUs (see \ref{subsec:gpu}) in SXM4 form factor are integrated in the blade.
The local memory subsystem of the GPU is placed in the same physical chip of the processing element, thus offering high density and performance.
This relies on the second generation High Bandwidth Memory express interface technology (HBM2e).
Each GPU has 64~GB of addressable memory that is organized into four 16~GB HBM2e stacks.
Each stack is controlled by two 512-bit memory controllers capable of 3200 MT/s.
Overall, more than a terabit per second can be delivered by each GPU, namely 1638 GB/s. 
In total, the local storage for GPU computation is 320 GB in capacity and can be accessed with an impressive 6.5 TB/s aggregated bandwidth.

\begin{figure}
    \centering
        \includegraphics[width=\linewidth]{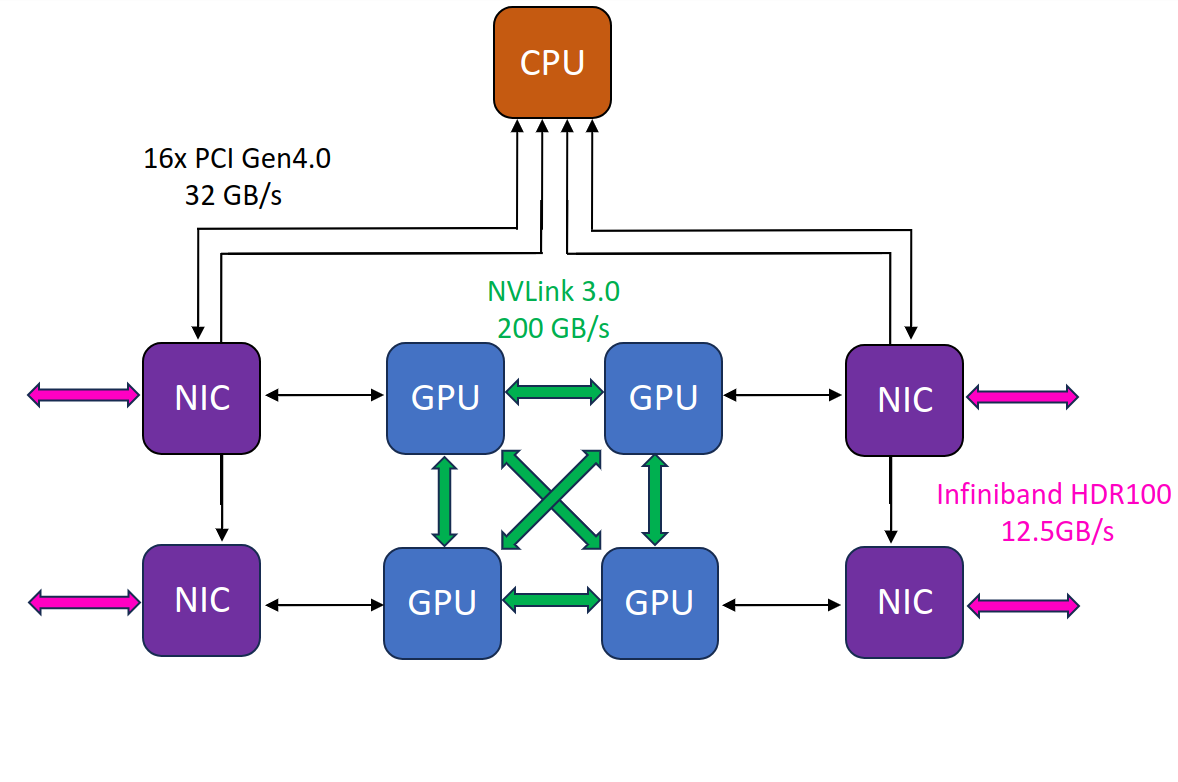}        
    \caption{Booster blade intra-node communication pattern (logic view)}
    \label{Fig:gpu_node}
\end{figure}

Intra-node communication pattern is depicted in Figure~\ref{Fig:gpu_node}.
The CPU utilizes four bundles of PCIe lanes to communicate independently with individual GPU.
A bundle consists of 16 PCIe Gen 4.0 lanes for a total of 32 GB/s bandwidth per CPU-GPU communication.
Total bandwidth available along the 64 lanes of the CPU is 128 GB/s.
Multi-GPU systems are supported by a proprietary fast high speed interconnect (\nvidia NVLink 3.0) that provides 200 GB/s bidirectional bandwidth per GPU pair, 600 GB/s in total.

The \gpuname blade is equipped with 2 dual-port Mellanox HDR100 ConnectX-6 \ib network interface cards (NIC) for inter-node communication. 
They provide an aggregated 400 \gbit bandwidth as well as CPU offloading features that are described in the next sections.

%% file: network.tex
\subsection{Network system} \label{sec:network}
\begin{figure} 
    \centering
    \includegraphics[width=0.7\textwidth]{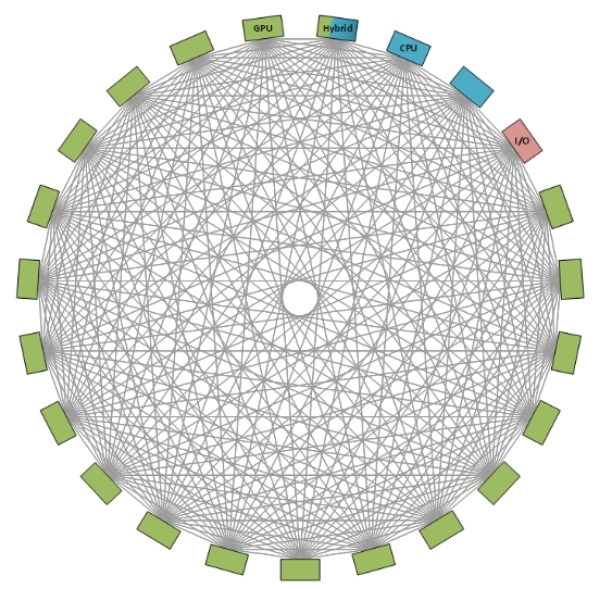}
    \caption{
        Internal network \emph{dragonfly} topology. 
        Colors indicate the technology of the underlying nodes.
        Green is used for Booster cells, blue for DC cells, pink for the I/O.
        See text for details.
        }
    \label{Fig:DragonFly}    
\end{figure}

The internal network of a cluster connects the compute nodes together.
\leo's network follows a scalable hierarchical cell-based architecture, with a cell being a collection of server nodes.
At top level, there are 23 cells fully connected in a \emph{dragonfly} topology~\citep{Ref:DF} as shown in Figure~\ref{Fig:DragonFly}.

Locally, intra-cell routers are organized in a bipartite graph in which a first tier is directly connected to servers (\emph{leaf routers}) and a second tier (\emph{spine routers}) is equally provisioned with down-links.
Such scheme, called \emph{dragonfly+}, allows twice the group size and  a factor four in scalability, when compared to the standard \emph{dragonfly} topology~\citep{Ref:DFplus}, it is denser and request less switches.

Nodes in the cluster are tighly coupled using 200~\gbit \ib Mellanox's High Data Rate (HDR) technologies components~\citep{Ref:HDR}.
The switch model is the QM8700, offering a latency of 90 nanoseconds port-to-port and up to 390 million messages delivery per second per port \citep{Ref:Switch}.
It can be used in two configurations, 40 ports at $200$~\gbit or 80 ports at $100$~\gbit bandwidth, with the latter widely adopted at leaf level and referred to as HDR100.
The total number of HDR switches is 823.

Spines and leafs have different arrangements, depending on the technology of the underlying node:
\begin{itemize}
\item the number of spine switches is 18 per cell regardless the cell type.
They are configured in 200G 40-port mode with 22 up-links and 18 down-links.
This corresponds to a pruning factor of 0.82 that is used to compensate for a $1.11$ near non-blocking factor of the leaf layer in the Booster cells.
\item Leaf switches organization depends on the cell type and is always HDR100, except the \emph{Fast Tier} where each links use full 200G HDR bandwidth per port (see Section \ref{sec:storage}).
Each node in the Booster partition is connected to two \emph{leaf} switches.
Differently, in the Data-Centric partition the nodes are directly connected to a single \emph{leaf} switch using a HDR100 link, i.e. 16 HDR ports on a switch serve 32 CPU nodes. 
For the hybrid cell, the two arrangements just described are combined, namely 6 out of 8 racks of the cell are DC style and the remaining 2 are Booster style.
The number of leaf switches per cell is 18 for Booster and Hybrid type and 16 for the DC type.
The I/O cell uses 13 leaf switches.
\end{itemize}
At node level, the network adapter is the ConnectX-6 card (CX6) which can sustain up 200 millions of messages per second with a latency of 600 ns~\citep{Ref:HDR_hca}.
The CX6 support PCIe Gen4 communication on 32 lanes including pass-through functionality.
Applications that do not require the entire bandwidth can benefit from the integrated PCIe switch that allows to serve up to 8 virtual machines on host\footnote{Known as MultiHost.}.
In addition, the CX6 comes with acceleration engines that provide CPU offloading for important HPC and AI tasks like: Remote Direct Memory Access (RDMA) for direct data movement from storage infrastructure to local GPU memory, transport operations like adaptive routing and congestion management, MPI collectives and tag matching, encryption based on personal user key.

Considering the latencies of the switch and of the NIC mentioned above and the following lengths of optical fiber - 1 meter from NIC to leaf, 5 meters from leaf to spine and 20 meters between the spines - the maximum latency between two nodes located at opposite side of the cluster is 3 microseconds.
In general, inter-node communication latency is dominated by the sending and receiving NICs that introduce 1.2 microseconds delay, independently from the destination.

Finally, four gateways routers are used to interface the cluster with external networks.
Each of these units provides eight 200~\gbit Ethernet-\ib protocol translators for a total bandwidth per unit of 1.6~Tbps and 6.4~Tbps~ aggregated~\citep{Ref:Gateway}.
In addition, an Ethernet administrative network is used for management, with dedicated switches at rack level and single port adapters on each node.

%% file: storage.tex
\subsection{Storage system} \label{sec:storage}
A 12-rack system provides storage functionality to the whole computer cluster.
The system is based on DDN's appliances and consists of two tiers to accommodate all requirements of modern HPC and AI diverse workloads .
\begin{itemize}
\item \emph{Fast Tier} provides $5.7$~PB of raw memory for IOPS eager applications and offers burst buffer capability for \emph{hot} data generally. 
It is composed by 31x ES400NVX2 appliances configured with $\simeq 150$~TB of solid state drives (SSD) using Non-Volatile Memory Express (NVMe) technologies.
\item \emph{Capacity Tier} is a $137.6$~PB raw capacity storage partition using SAS Hard Disk Drive (HDD) components.
It consists of 31 modules composed by a controller head (ES7990X) and two SAS expansion enclosures (SS9012)  housing a total of 246 by 18~TB HDD, providing 4.4 PB of storage capacity per module.
Metadata is handled by four additional flash-based ES400NVX units.
\end{itemize}

Overall, the storage system consists of 66 DDN's appliances together with the related software stack that is essential for high speed data movement in the different computing scenarios that \leo can serve, in addition to standard fault tolerance and security functions.
Table \ref{Tab:filesystem} shows the mapping of the three global namespaces to the hardware resources just described, together with the related size and bandwidth characteristics.

The filesystem is based on Lustre~\citep{Ref:Lustre} and supports encryption and multi-tenancy.
The first is a feature for security and isolation that allows to access selected portion of the storage namespace to authenticated users only.
This is based on CryptoFS \citep{Ref:CryptoFS}. 
The second  feature allows multiple access (multiple client) to files.
Of prominent importance for AI workloads, \emph{GPUDirect} technology  is also supported by the storage system,
it can directly use the GPU memory for I/O, avoiding the use of system memory (RAM) as bounce buffer.
With objects striped across multiple disks, Lustre provides also parallel access to large files at near-wire speed.

\begin{table} [b]
\centering
\begin{tabular}{|c|c|c|c|c|c|}
\hline
Work area                 & ES7990X & ES400NVX2 & ES400NV  & NetSize & Bandwidth \T\\
                          &    \#    &     \#     &    \#     &   PiB    &  GB/s\B\\
\hline
\texttt{\slash home}      & -      &    4      & -        &    0.5  &  240\T \\ 
\texttt{\slash archive}   & 18     &     -     & 2        &    53.9  &  360\\ 
\texttt{\slash scratch}   & 13     &    27     & 2        &    42.4  & 1300\B\\ 
\hline
\end{tabular}
\caption{Filesystem organization and specifications}
\label{Tab:filesystem}
\end{table}

%% file: service.tex
\subsection{Frontend and service partitions} \label{sec:service}
The Frontend partition provides the user with access to the system.
Typical operations on frontend nodes encompass software development, code compilation, data management, interface to other systems on site, job submission, data pre-processing, data post-processing and results visualization.
In \leo the frontend nodes see both compute partitions and the global filesystems as well.
The number of frontend servers is 32, each equipped with dual socket Xeon Scalable processor (32 cores, \intel 3rd Gen, the same model of the Booster's compute node) and 16 DDR4-3200 channels per socket.
Sixteen nodes are dedicated to login and are configured with a local 6 TB disk space in RAID-1 configuration.
The other 16 nodes are specialized for post-processing visualization and are equipped with NVMe disks ($6.4$~TB total local capacity) and 2 NVIDIA PCIe Quadro RTX8000 by $48$~GB RAM each.

In order to deploy, manage and monitor the \leo cluster that accounts for about 5000 compute nodes, 11 tailored servers are used, called Operational Managment Nodes (OMN).
OMNs feature a single AMD EPYC \emph{Rome} CPU with 64 cores and 3 three dual ports NICs supporting 10 GbE, 50 GbE and HDR100.

The complete list of hardware components is reported in Appendix~\ref{sec:hardware}.

%% file: software.tex
\subsection{Software ecosystem} \label{L_SW}
\leo runs Red Hat Enterprise Linux 8 operating system on all nodes and uses SLURM as workload manager~\citep{ref:slurm}.
Two architecture-specific suites are installed, namely \intel OneAPI, \nvidia HPC SDK.
The latter includes a complete software stack to build AI applications with highly optimized libraries such as cuDNN for deep neural networks and NCCL for multi-GPU communication.
The GNU compiler collection is also installed.

Software management is done using Spack~\citep{ref:Spack} and Environment Modules~\citep{ref:module}.
A large set of HPC programming tools is available for developers, based both on closed and open source products.
The software for scientific production is organized on a category basis, serving each research community with dedicated pre-installed tools
e.g. chemestry-physics, deep learning, life sciences and meteo.
Further details and updates can be found in the user guide available on the website of CINECA \citep{ref:userguide}.
Baseline tools are listed below.

\begin{itemize}

    \item Parallel profilers and debuggers
    \begin{itemize}
        \item GNU debugger (GDB) 
        \item \intel debugger (IDB) and VTune profiler
        \item \nvidia Nsight profiler (System and Compute) and CUDA-GDB
        \item Valgrind 
    \end{itemize}

    \item Communication libraries
    \begin{itemize}
        \item OpenMPI, 
        \item \intel MPI
    \end{itemize}

    \item Numerical application libraries
    \begin{itemize}
        \item \intel Math Kernel Library, 
        \item GNU scientific library, 
        \item Math and Python libraries
    \end{itemize}

    \item Containerization is supported through several different tools:
    \begin{itemize}
        \item Syslab \href{https://sylabs.io/}{Singularity} Enterprise edition
        \item \nvidia Container Framework and \href{https://github.com/NVIDIA/pyxis}{Pyxis} Slurm plugin  
	\item ParTec Parastation also supports the execution of containerized applications, improving the flexibility of a pure Singularity approach;
    \end{itemize}

    \item Monitoring is operated via Atos SMC xScale suite, based on Prometheus, and using Grafana as frontend. 
          Detection and tracking of issues is performed by Parastation HealthChecker.

\end{itemize}

%% file: power.tex
\subsection{Power consumption, cooling and management} \label{L_power}

\leo is hosted by CINECA in its new data center at the Big Data Technopole~\citep{Ref:tecnolopolo} in Bologna, Italy.
The room floor has been designed with a plan in two steps to support the current pre-exascale and a future exascale machine.
Presently, the data center features 10~MW of IT load with $ 1240~m^2 $ of computing floor space and $ 900~m^2 $ of ancillary space.
The second step considers an increased power support up to $20$ MW IT load and $2600~m^2$ additional computing floor.

All major components of \leo are cooled down using warm water-cooling technology, including the power supplies.
The inlet water temperature is 37 Celsius degrees and the total Direct Liquid Cooling capacity is 8 MW.
The system is pretty efficient with a 1.1 value of Power Usage Effectiveness (PUE).
This means that the overhead needed to cool down \leo is the $10\%$ of the power used to feed it.

Energy consumption of the cluster is controlled by means of various tools including two ATOS proprietary software products (Bull Energy Optimizer and Bull Dynamic Power Optimizer).
One allows to log time profiles of energy and temperature via IPMI and SNMP protocols and to cap the clock frequency of the CPUs depending on the total power consumption.
The other is used to find the best workpoint in terms of energy consumption and performance of a running application i.e. reducing the power absorption by adjusting the clock frequency with limited performance degradation.
Concerning GPU, a vendor specific manager tool (NVIDIA Data Center GPU Manage) is used to limit device clock when a configurable energy threshold is surpassed.

%% file: outro.tex
\section{Access to \leo}

\leo is a EuroHPC-JU system that is hosted and operated by CINECA supercomputing center.
Researchers from academia, research institutes, public authorities, and industry can apply for access to computing time.
The access is mainly based on \emph{Calls for Proposal} from EuroHPC ($50\%$) and CINECA ($50\%$) via its ISCRA program (Italian SuperComputing Resource Allocation).
Submitted proposals are peer reviewed for scientific merit and undergo a technical assessment for suitability to perform on \leo architectures, in order to ensure the highest scientific reach of the selected project.
Detailed information are available at \citep{ref:Access} and \citep{ref:ISCRA} webpages.

\subsection*{Acknowledgement}
The acquisition and operation of \leo supercomputer is funded jointly by the Italian Ministry of University and Research and by the EuroHPC Joint Undertaking (EuroHPC JU) under grant agreement \emph{N. cnect.ddg1.c.2(2019)8804531 - \leo supercomputer} through the European Union's Connecting Europe Facility and the Horizon 2020 research and innovation programme.
The EuroHPC JU is a legal and funding entity created in 2018 to enable the European Union and EuroHPC participating countries to coordinate their efforts and pool their resources with the objective of making Europe a world leader in supercomputing.

The authors thank ATOS for the solution provided and all the key technology partners \nvidia, \intel, DDN, for their support during design, construction, delivery and testing.

%% file: benchmarks.tex
\section{Benchmark results} \label{appendix:benchmark}

The procurement process of \leo relied on the evaluation of prominent and application-specific workloads.
Here, results for the following benchmarks are sketched as a list of tables.
\begin{itemize}
    \item Synthetic HPC benchmarks
    \begin{itemize}
        \item High Performance Linpack (HPL) 
        \item High Performance Conjugate Gradients (HPCG)  
        \item IO500
    \end{itemize}
    \item Application benchmarks
    \begin{itemize}
        \item \href{https://www.quantum-espresso.org/}{QuantumEspresso}:  electronic-structure calculations and materials modeling
        \item \href{https://github.com/SPECFEM/specfem3d_globe}{SPECFEM3D Globe}: global seismic wave propagation
        \item \href{http://plutocode.ph.unito.it/}{PLUTO}: astrophysical gas dynamics  
        \item \href{https://github.com/milc-qcd/milc_qcd}{MILC}: lattice QCD calculations 
        \item Lattice Boltzmann Method (LBM):  computational fluid dynamics 
    \end{itemize}
\end{itemize}

\subsection{HPL, HPCG and Green500}
Table~\ref{Tab:LeonardoS} summarize the metrics of \leo Booster as presented at TOP500 in June 2023~\citep{Ref:Top500_2306}.
The most relevant is a $238.7$~\pf measured HPL performance out of 304.5 \pf of theoretical peak.
Such a result was achieved by using 3300 compute nodes.
Total power consumption was 7.4 MegaWatts, thus giving an average performance per unit of power of $32.2$ \gf/W.
These results earned \leo the rank $4^{th}$ and $15^{th}$ in the TOP500 list and Green500 list respectively.
In the same edition, \leo was also ranked 4th in the HPGC category with a performance of $3.11$~\pf.

\begin{table} [h] 
\centering
\begin{tabular}{|c|c|c|} \hline
Benchmark  & Performance [\pf]   &  Rank\T \B   \\ \hline 
HPL        & 238.7               &  4\T\\
HPCG       &  3.11               &  4\T\B    \\ \hline
\end{tabular}
\caption{\leo performance at TOP500 in June 2023.}
\label{Tab:LeonardoS}
\end{table}

\subsection{IO500}
At ISC 2023 among the production machines, \leo was $1^{st}$ in the bandwidth category of IO500 list~\citep{Ref:IO500}.
Related performances figures are shown in Table~\ref{Tab:IOperformance}.
Standard \verb|ior| benchmark results are 1533 GiB/s and 1883 GiB/s bandwidth for \verb|ior-easy-write| and \verb|ior-easy-read| respectively.

\begin{table} [h] 
    \centering
    \begin{tabular}{|c|c|c|c|c|}\hline
      Benchmark  & Score  & BW (GiB/s) & MD (KIOP/s) & Rank\T\B \\ \hline
    IO500   & 649  & 807   & 522         & 1\T\B            \\ \hline  
    \end{tabular}
    \caption{\leo IO500 performance at ISC2023.}
\label{Tab:IOperformance}
\end{table}

\newpage
\subsection{Application Benchmarks}

Table \ref{Tab:ApplicationBenchmarks} shows the results of domain-specific application benchmark in terms of \emph{Time-to-Solution} (TTS) in seconds and \emph{Energy-to-Solution} (ETS) in kWh.
The job size in terms of number of nodes spans from 12 to 32.  
In case of PLUTO  the ETS has been estimated using CPU power consumption only, since the program does not use GPUs.

\begin{table}  [h] 
    \centering
    \begin{tabular}{|c|c|c|c|c|}\hline
Application name & Domain              & Nodes & TTS   & ETS \T\B \\ \hline
QuantumEspresso  & Quantum Chemistry       &  12   & 439   & 1.14\T  \\
MILC	         & Quantum Chromodynamics  &  12   & 178   & 0.56 \\
SPECFEM3D        & Solid Earth             &  16   & 270   & 1.43\\
PLUTO	         & Astrophysics            &  32   & 2874  & 11.7\B\\ \hline
    
    \end{tabular}
    \caption{\leo application benchmarks performance.}
\label{Tab:ApplicationBenchmarks}
\end{table}

\begin{figure} [h]
    \centering
     \includegraphics[width=0.5\linewidth]{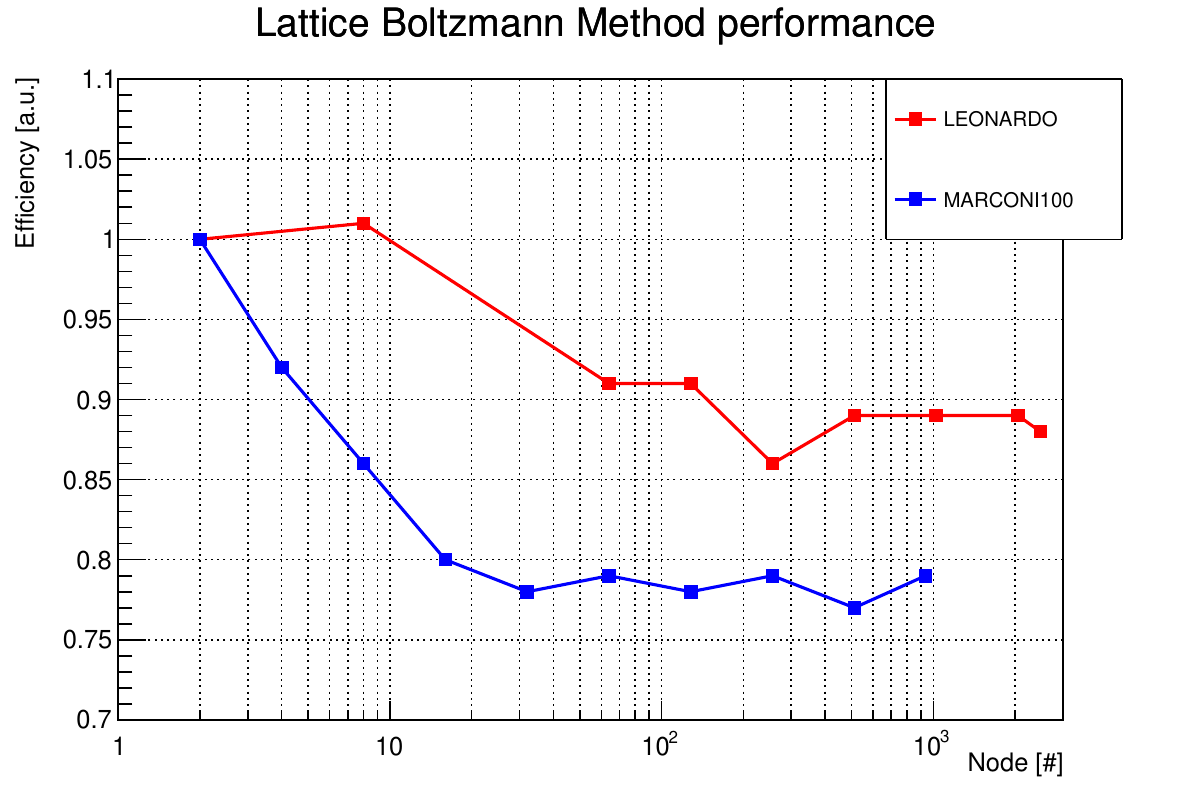}    
         \caption{LBM Weak scaling efficiency comparison}
         \label{fig:lbm}
\end{figure}

Figure \ref{fig:lbm} presents the weak scaling of the Lattice Boltzmann Method (LBM) benchmark, a code that has been described in details by \cite{ref:Nature} and \cite{ref:EXAlb}. 
Considering the scaling efficiency of the same code on \emph{Marconi100}, a CINECA's GPU based cluster equipped with \nvidia V100, a significant better performance has been measured.
In terms of TTS, \leo was about $2.5$ times faster than \emph{Marconi100} as documented by \cite{ref:LBM_m100}.
Table \ref{Tab:LBM} presents the performance of \leo in terms of Lattice Updates per Seconds (LUPS) and explicits the total number of GPUs in each point of the scaling.

\begin{table}  [h] 
    \centering
        \begin{tabular}{|c|c|c|c|}\hline
            Nodes [$\#$] &  \#GPUs &            Performance [LUPS $\times 10^{12}$] & Efficiency\T \B\\ \hline
               2  &    8  &  0.0476 & 1.00\T \\
               8  &   32  &  0.192  & 1.01 \\
              64  &  256  &  1.38   & 0.91 \\
             128  &  512  &  2.76   & 0.91 \\
             256  &  1024 &  5.24   & 0.86 \\
             512  &  2048 & 10.8    & 0.89 \\
            1024  &  4096 & 21.6    & 0.89 \\
            2048  &  8196 & 43.3    & 0.89 \\
            2475  &  9900 & 51.2    & 0.88\B \\\hline
        \end{tabular}
    \caption{LBM Weak scaling efficiency }
\label{Tab:LBM}
\end{table}





%% file: hardware.tex
\newpage
\section{Hardware components} \label{sec:hardware}
As of this writing, LEONARDO supercomputer consists in the following components.

\paragraph{Booster partition}
\begin{itemize}
    \item 3456 nodes (13824 GPUs)
    \item single node \emph{Da Vinci} blade, based on the BullSequana X2135 
    \begin{itemize}
        \item 1x 32-core \intel Xeon Platinum 8358 CPU, 2.6~GHz (\emph{Icelake})
        \item 8x 64~GB DDR4-3200 (512~GB) 
        \item 4x \nvidia custom Ampere A100 GPU 64 GB HBM2
        \item 2x dual-port HDR network interface (400 \gbit aggregated)
    \end{itemize} 
\end{itemize}

\paragraph{Data-Centric partition}
\begin{itemize}
    \item 1536 nodes (172032 CPU cores) 
    \item three-node BullSequana X2140 blade, each node with    
    \begin{itemize}
        \item 2x 56-core \intel Xeon Platinum $8480$\texttt{+} CPU, 2.0~GHz (\emph{SapphireRapids})  
        \item 16x 32~GB DDR5-4800 (512~GB)
        \item 1x SSD 3.84~TB M.2 NVMe
        \item 1x single port HDR100 network interface (100 \gbit)        
    \end{itemize}
\end{itemize}

\paragraph{Storage}
\begin{itemize}
    \item Fast Tier, 5.7~PB full flash 
    \begin{itemize}
        \item 31x DDN appliance ES400NVX2 configured with
        \begin{itemize}
            \item 24x SSD 7.68~TB NVMe with encryption support ($184.3$~TB)
            \item 4x \ib HDR ports (800 \gbit aggregated)
            \item metadata resource included
        \end{itemize}  
    \end{itemize}

    \item Capacity Tier, $137.6$~PB
     \begin{itemize}
        \item 31x DDN appliance ES7990X configured with  
        \begin{itemize}
            \item 1 Controller head (82 disks) + 2 expansion enclosures (SS9012, 164 disks)
            \item 246x HDD 18~TB SAS 7200 rpm (4.4~PB) 
            \item 4x \ib HDR100 ports (400 \gbit aggregated)
        \end{itemize}
        \item 4x DDN appliance SFA400NVX for metadata (322~TB), configured with
        \begin{itemize}
            \item 21x SSD 3.84~TB NVMe with encryption support ($80.8$~TB)
            \item 8x \ib HDR100 ports (800 \gbit aggregated)
        \end{itemize}
    \end{itemize} 
\end{itemize}

\paragraph{Frontend partition}
\begin{itemize}
    \item 32 Frontend nodes (16 login + 16 graphical)
    \begin{itemize}
        \item 2x 32-core \intel Xeon Platinum 8358 CPU, 2.4~GHz (\emph{Icelake})
        \item 16x 32~GB DDR4-3200 (512~GB)
        \item 1x HDR100 network interface
        \item 2x 50~GbE network interface.
        \item \emph{Login}, BullSequana X430-E6
        \begin{itemize}
            \item 6 TB HDD in RAID1
        \end{itemize}
        \item \emph{Graphical}, BullSequana X450-E6
        \begin{itemize}
            \item 6.4 TB NVMe
            \item 2x GPU PCIe NVidia Quadro RTX8000 48 GB
        \end{itemize} 
    \end{itemize}

\end{itemize}

\paragraph{Service partition}
\begin{itemize}

    \item 11 Operational Management nodes (3 Master + 8 Worker)
    \begin{itemize}
        \item 1x 64-core CPU AMD EPYC 7h12 (Rome), 2.6 GHz, TDP 280 W
        \item 1x dual-port HDR100 network interface
        \item 1x dual-port 50~GbE network interface
        \item 1x dual-port 10~GbE network interface
        \item \emph{Master}
        \begin{itemize}  
        \item 8x 16~GB DDR4-3200 (128~GB) 
            \item 2x 960~GB NVMe with M.2 slots
            \item 2x 3.84~TB 2.5 inches SATA3 SSD
        \end{itemize} 
        \item \emph{Worker} 
        \begin{itemize} 
        \item 16x 32~GB  DDR4-3200 (512~GB) 
            \item 2x 3.2~TB NVMe U.2
            \item 4x 3.84~TB 2.5 inches SATA3 SSD
            \item 8x 12~TB 3.5 inches SATA3 HDD
        \end{itemize} 
    \end{itemize}

\end{itemize}